\documentclass{amsart}
\usepackage{graphicx}

\title{Low temperature microwave emission from molecular
clusters}
\author{A. Hern\'{a}ndez-M\'{\i}nguez \and M. Jordi \and R Amig\'{o}
\and A. Garc\'{\i}a-Santiago \and J. M. Hernandez \and J. Tejada}
\dedicatory{Dept. de F\'{\i}sica Fonamental, Universitat de
Barcelona,\\Avda. Diagonal 647, 08028 Barcelona, Spain}

\begin{document}

\begin{abstract}
We investigate the experimental detection of the electromagnetic
radiation generated in the fast magnetization reversal in
Mn$_{12}$--acetate at low temperatures. In our experiments we used
large single crystals and assemblies of several small single
crystals of Mn$_{12}$--acetate placed inside a cylindrical
stainless steel waveguide in which an InSb hot electron device was
also placed to detect the radiation. All this was set inside a
SQUID magnetometer that allowed to change the magnetic field and
measure the magnetic moment and the temperature of the sample as
the InSb detected simultaneously the radiation emitted from the
molecular magnets. Our data show a sequential process in which the
fast inversion of the magnetic moment first occurs, then the
radiation is detected by the InSb device, and finally the
temperature of the sample increases during 15 ms to subsequently
recover its original value in several hundreds of milliseconds.
\end{abstract}

\maketitle

Molecular clusters are nanomagnets showing important phenomena
associated to their magnetic anisotropy and the possibility to
tune their quantum mechanical properties by applying a magnetic
field. The discovery of the resonant spin tunneling between the
degenerate spin levels at both sides of the anisotropy energy
barrier \cite{1,2} was the first sign that quantum mechanics can
reveal itself in these magnetic units made of several hundreds of
atoms. Since then, near thousand papers on molecular magnets have
been published \cite{3}. Very recently a new field combining
molecular magnets, magnetization studies, cavities and
electromagnetic radiation has attracted the attention of different
groups. The most recent works on this particular topic deal with
the absorption and emission of microwaves
\cite{4,5,6,7,8,9,10,11,12,13,14,15}. In this letter we report low
temperature experimental studies of the detection of microwave
emission of Mn$_{12}$ single crystals inside a cylindrical
waveguide. This work follows that published very recently
\cite{13}, focusing now on the time sequence of the reversal of
the magnetic moment, the emission of radiation and the heat
release in the sample on the millisecond scale.

\begin{figure}
\includegraphics[width=8cm]{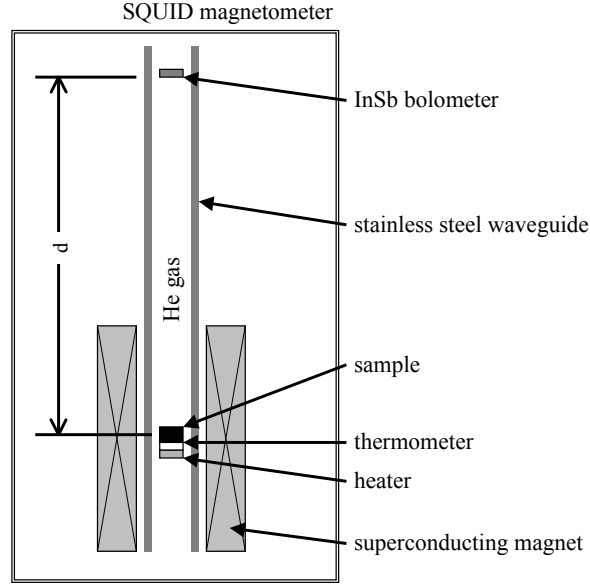} \caption{Experimental setup for the
detection of radiation of the Mn$_{12}$ sample.} \label{1}
\end{figure}

In Figure~\ref{1} we show the details of our experimental setup.
Two kinds of samples were investigated. In some experiments, the
sample was made by assembling together several tens of oriented
single crystals of Mn$_{12}$--acetate with a total volume around
20 mm$^3$.  In some others, we used only one large single crystal
having a magnetic moment $M$ close to one third of that of the
assemblies of crystals ($M \sim$ 1 emu). An InSb hot electron
device, with a spectral bandwidth of 60 GHz -- 3 THz, was used to
detect the electromagnetic radiation possibly emitted by the
sample. Both the sample and the detector were placed inside a
stainless steel cylindrical waveguide and the whole assembly was
inserted in a commercial rf-SQUID (Superconducting QUantum
Interference Device) magnetometer that allowed to change the
temperature, $T$, and the magnetic field intensity, $H$, of the
experiment. The distance, $d$, between the sample and the detector
ranged from 8 to 20 cm.

In our experiments we measured simultaneously the magnetic moment
and the temperature of the sample, $M$ and $T_{\rm{s}}$, and the
voltage drop at the detector, $V_{InSb}$. $M$ was measured using
either the rf-SQUID magnetometer, with a time resolution of 20 s,
or a 30-turn, 4-mm-diameter copper detection coil, with a time
resolution of 10 $\mu$s. $T_{\rm{s}}$ was measured by an attached
ruthenium oxide thermometer with a time resolution of 2 ms at $T
=$ 4.2 K. $V_{InSb}$ was measured as a function of temperature and
magnetic field. The experimental resolution time for the voltage
drop at the InSb device was about 1 ms. The thermometer and the
detector were calibrated at the different temperature and magnetic
field values of the experiment. The easy axis of the crystals and
the applied magnetic field were checked to be parallel to the axis
of the waveguide and the magnetic properties of different sets of
crystals were measured with the rf-SQUID magnetometer.

We swept the applied magnetic field up and down between -2 T and 2
T with a sweeping rate d$H$/d$t =$ 30 mT/s for different
experimental temperatures. Prior to all these experiments, we
characterized both crystallographically and magnetically (we
performed dc and ac measurements to get the resonant field values
and the blocking temperature) all the crystals used in our
experiments to confirm that the materials were only made of pure
single Mn$_{12}$--acetate phase. The most important result is that
the total magnetic moment reverses orientation at one of the
different resonant fields, depending on temperature. That is,
instead of observing only regular steps associated to resonant
spin tunneling, single large jumps are seen at the different
resonances. For instance, for the experiments carried out at $T =$
2.0 K, the inversion of the total magnetic moment $M$ occurs at $H
=$ 1.38 T,  while at  $T =$ 2.6 K it does occur at $H =$ 0.98 T.

\begin{figure}
\includegraphics[width=8cm]{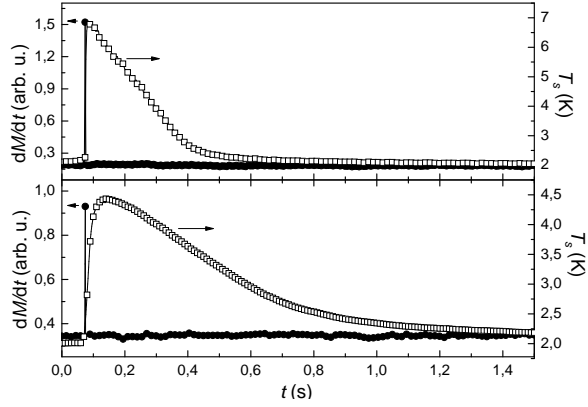} \caption{Temporal evolution of
$T_{\rm{s}}$ (right scale) and d$M/$d$t$ (left scale) around the
magnetic field value at which the magnetization reversal occurs,
for an individual single crystal (upper panel) and an assembly of
single crystals (lower panel) ($T =$ 2.0 K).} \label{2}
\end{figure}

\begin{figure}
\includegraphics[width=8cm]{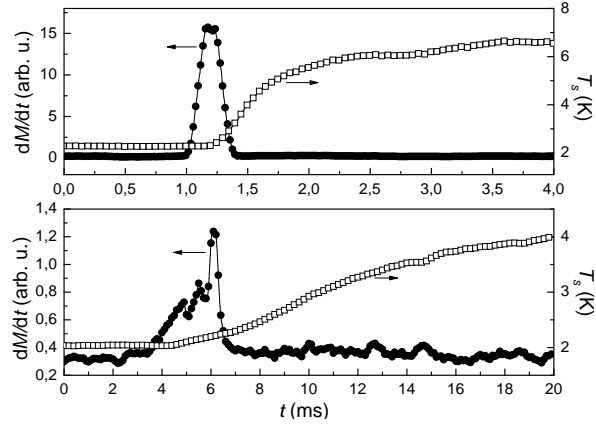} \caption{Magnification at short
times of the temporal dependence of $T_{\rm{s}}$ (right scale) and
d$M/$d$t$ (left scale) around the magnetic field value at which
the magnetization reversal occurs, for an individual single
crystal (upper panel) and an assembly of single crystals (lower
panel) ($T =$ 2.0 K).} \label{3}
\end{figure}

\begin{figure}
\includegraphics[width=8cm]{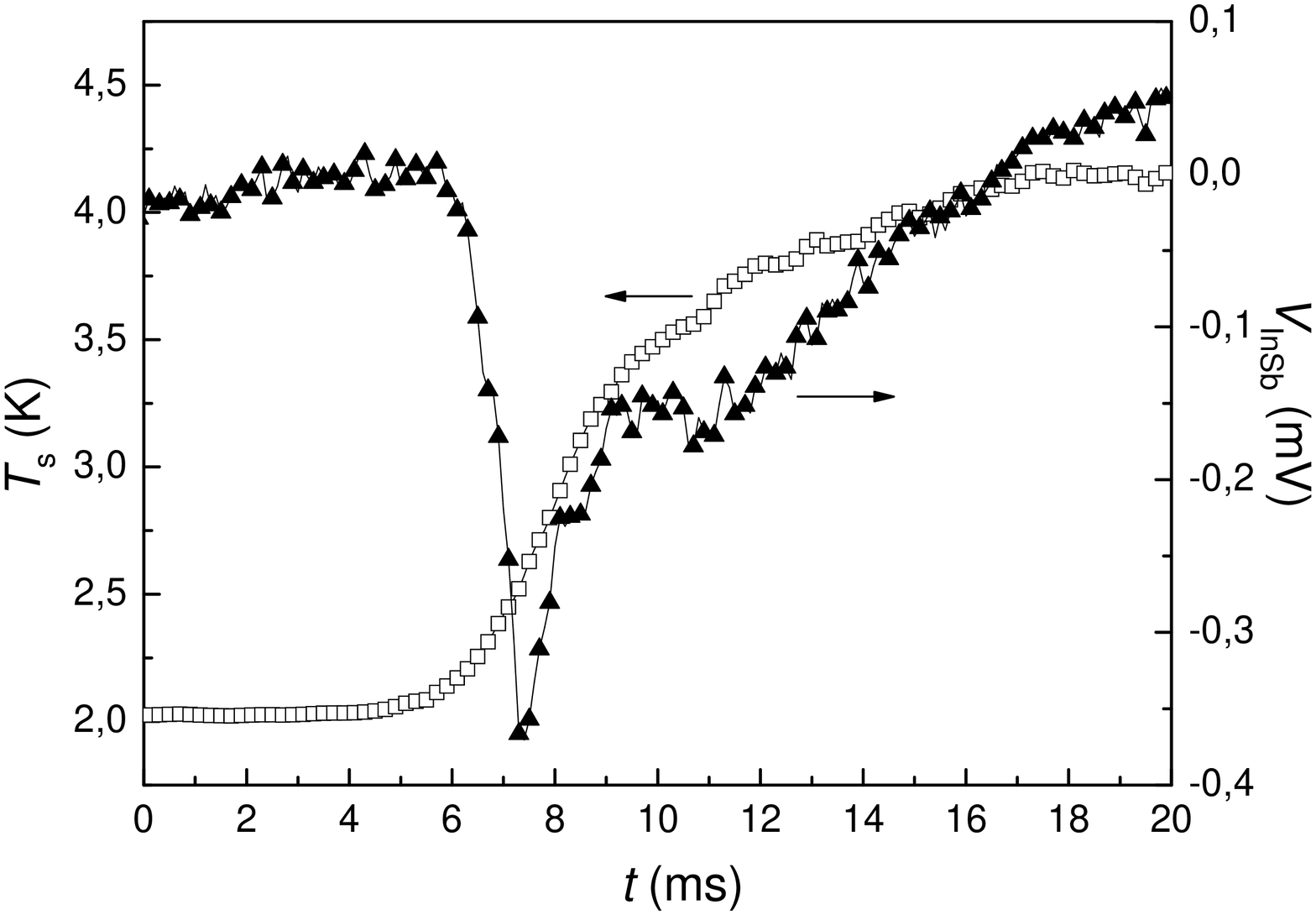} \caption{Time variation of
$V_{InSb}$ (right scale) and $T_{\rm{s}}$ (left scale) due to the
magnetization reversal at $T =$ 2.0 K for an assembly of single
crystals.} \label{4}
\end{figure}

\begin{figure}
\includegraphics[width=8cm]{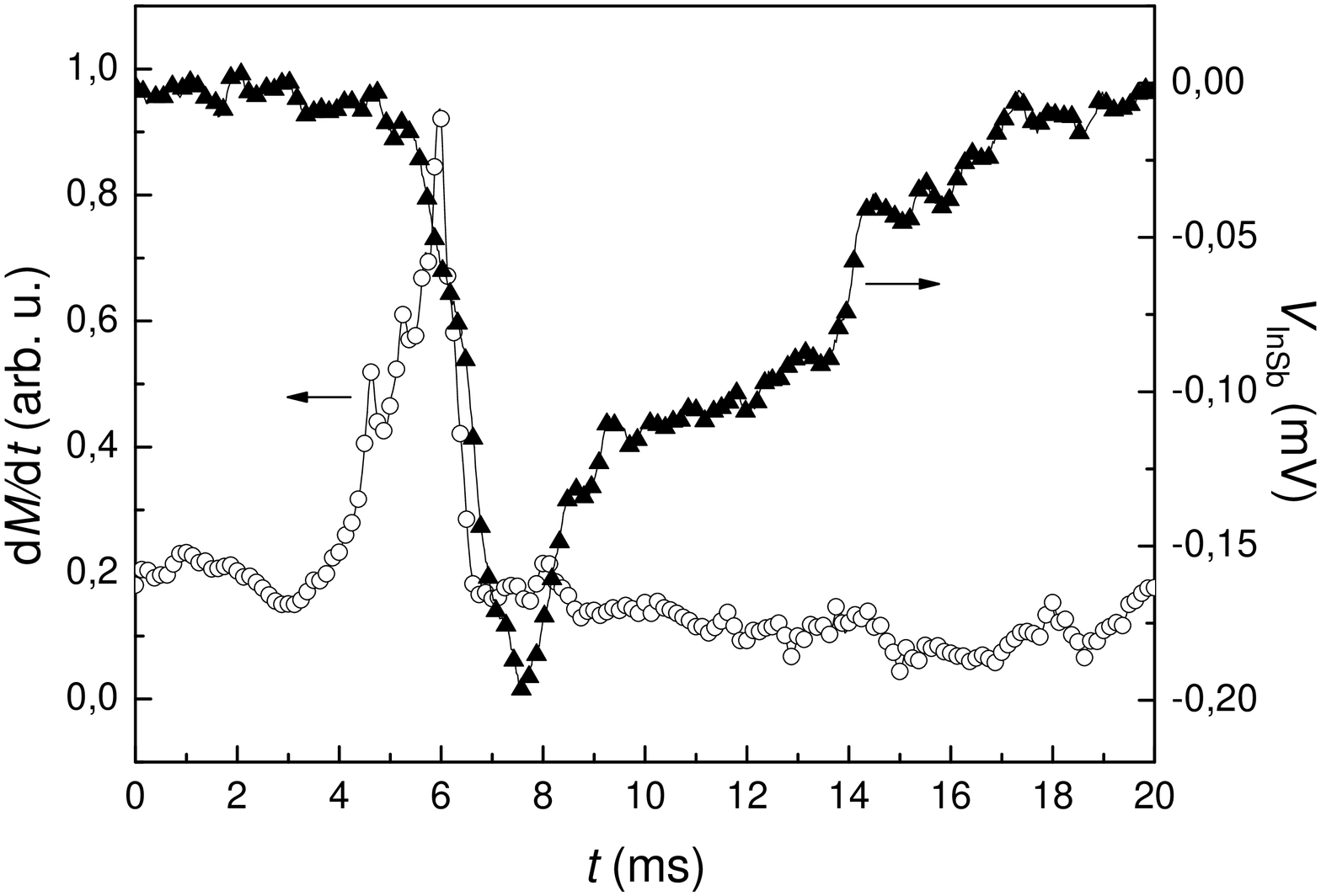} \caption{Time variation of
$V_{InSb}$ (right scale) and d$M/$d$t$ (left scale) due to the
magnetization reversal at $T =$ 2.4 K for an assembly of single
crystals.} \label{5}
\end{figure}

In Figure~\ref{2} we show the time dependence of $T_{\rm{s}}$ and
the time derivative of $M$, d$M$/d$t$, around the field value at
which the magnetization reversal occurs for the cases of using a
single crystal (upper panel) and an assembly of single crystals
(lower panel). A short time magnification of Figure~\ref{2} is
shown in Figure~\ref{3}. For the case of the large single crystal,
only one well defined magnetization reversal occurs (upper panel),
while several independent and sequential magnetization reversal
are present in the case of the assembly of single crystals (lower
panel). It is worth mentioning that in all cases the complete
reversal of the total magnetic moment takes place in less than 3
ms. In fact, it takes longer when the temperature increases ({\it
e.g.}, 0.8 ms at $T =$ 1.8 K and 2 ms at $T =$ 2.6 K, for the case
of the single crystal). The maximum variation of the temperature
takes place in a time of a few milliseconds and a subsequent
thermal relaxation of several hundreds of milliseconds follows to
recover the initial temperature. The most interesting and
intriguing results here are two: $T_{\rm s}$ reaches a maximum
value several milliseconds after the magnetization reversal
occurs, and this value never goes above 4 K for the assembly of
crystals and about 7 K for the large single crystal.

Figures~\ref{4} and~\ref{5} present the time dependence of the ac
voltage signal of the detector, $V_{InSb}$, superimposed
respectively to the time evolution of $T_{\rm s}$ and d$M$/d$t$,
for the cases of the experiments carried out at $T =$ 2.0 and 2.4
K using an assembly of single crystals. The voltage drop observed
in the figures is made first of a fast signal of a few
milliseconds duration and takes around 15 ms to recover its
initial value. As in the previous figures, the initial time ($t =$
0) corresponds to a field value very close (the deviation is of
the order of 0.01 Oe) to the magnetic field at which the first
magnetisation reversal takes place. These phenomena are not
detected above $T =$ 3.0 K.

We also verified that warming the sample up to near 20 K by
electric pulses of 100 ms through a ruthenium oxide heater in
contact with the Mn$_{12}$ sample produces a single voltage drop
at the detector of the same intensity and a time duration of
hundreds of milliseconds. The voltage drop in this case is then
different from the case of the magnetization reversal, as the
detector does not register any fast pulse with a duration of a few
milliseconds.

The Mn$_{12}$ molecules have total spin $S = 10$ and their
magnetic properties are described by the spin Hamiltonian

\begin{equation}\label{eq1}
{\mathcal H} = -D S_z^2 - F S_z^4 - g \mu_B H_z + {\mathcal H}',
\end{equation}

\noindent where $D = 0.55$ K, $F = 1.2\times 10^{-3}$ K, $g =
1.94$ and ${\mathcal H}'$ contains small terms that do not commute
with $S_z$. The two energy minima corresponding to $m = \pm 10$
are separated by the so-called magnetic anisotropy barrier height
$U = 65$ K. The $m$ and $m'$ eigenvalues of $S_z$ come to
resonance at the field values

\begin{equation}
\label{eq2} H_{m,m'} =n\frac{D}{g\mu _B }\left[ {1+\frac{F}{D}\left(
{m^2+{m'}^2} \right)} \right],
\end{equation}

\noindent where $m + m' = -n$. At $H_z  \neq  H_{m,m'}$,
transitions between positive and negative $m$ occur due to thermal
activation over the barrier height. However, at the resonance
fields, $H_z = H_{m,m'}$, the transitions are combinations of
thermal activation and quantum tunneling. At each resonant field
the transition is mostly dominated by a pair $(m, m')$, which
changes with the temperature of the sample. At the microscopic
level, a Mn$_{12}$ molecule must be thus activated from the $m =
-10$ level to the lowest $m$ from which tunnel becomes significant
on the time scale of the experiment. The spin of the molecule then
tunnels across the energy barrier and relaxes down the spin level
staircase to the $m' = 10$ state, that is, towards the direction
of the magnetic field. The corresponding energy difference between
the $m = -10$ and the $m' = 10$ state is released in the form of
phonons and photons. The phonon emission for the spin transition
in one particular molecule is however much more probable than the
corresponding photon emission. It has also been suggested that the
total magnetization reversal due to spin-phonon avalanches is
produced when the energy of the phonons released by many spin
transitions cannot thermalise and cause the sample to warm up
\cite{16,17,18,19}.

The maximum total Zeeman magnetic energy, $E_{\rm{mag}} \simeq
2M_{\rm{sat}}H$, released is of the order of mJ. As stated above,
the heating associated to the avalanche increases the temperature
of the sample to circa, but not above, 4 K in the case of
assemblies of single crystals and about 7 K when using only one
large single crystal. This experimental fact has been verified in
all the experiments carried out at different temperatures and with
total inversion of the magnetic moment occurring at different
magnetic fields. Most important, though, is the fact that the
sample starts heating once the total magnetic moment has reversed
orientation. Moreover, the radiation detection overlaps with the
reversal of the magnetic moment and occurs also before the
temperature of the sample reaches its maximum value. We should
however take these experimental facts with precaution, because the
resolution time of the thermometer is of the order of several
milliseconds. In any case, these data may also suggest that
thermal avalanches are not necessary to produce the total
inversion of the magnetic moment and the radiation emitted from
the sample is coupled mostly to the very fast magnetic reversal.

The fact that the magnetic reversal is accomplished first in a
time of less than 3 ms at $T =$ 2.0 K may be an indication that
the tunneling process occurs from the ground state ($m = -$10).
This would open the possibility of having the phonon laser effect,
suggested very recently by Chudnovsky and Garanin \cite{20}, as
the mechanism inducing spin tunnelling. The inversion of the total
magnetic moment of the sample consists in this case in the
coherent fast reversal of the magnetic moment of many molecules by
tunnelling caused by phonon vibrations. As a consequence, a huge
number of molecules is driven to one of the excited states $m'$ of
the right side of the potential well before the occurrence of the
staircase decay to the ground state $m' =$ 10.

We come now to the point of how to explain the detection of
radiation by the InSb in the absence of enough thermal black body
radiation emitted by the set of crystals. For an individual
molecule, the rate of spontaneous decay from the $m$ level to the
$m+1$ level due to photons is much lower than that for phonons,
$\Gamma_{m}^{\rm{photons}}/ \Gamma_{m}^{\rm{phonons}} = (v/c)^3$,
where $v$ and $c$ are respectively the speeds of sound and light
in the material, as a consequence of their very different density
of states \cite{21}. The detection of electromagnetic radiation
imposes, therefore, the following question: under which conditions
may we expect to detect photons rather than phonons?

A plausible explanation is that the detected electromagnetic
radiation could be superradiance. In his seminal paper published
in 1954 \cite{22,23,24}, P. H. Dicke first postulated that in the
case of having a set of $N$ quantum dipoles of the size of the
wavelength of the emitted photons, a spontaneous phase locking of
the quantum dipoles through the medium should occur, and a short
burst of radiation, the so-called superradiance, should be
emitted. Superradiance emission from molecular magnets has also
been recently suggested theoretically as a result of the fast
level crossing \cite{12}. In our experiments, the wavelength of
the emitted photons between the spin states $m$ and $m + 1$ and
the size of the set of magnetic quantum dipoles are both of the
order of a few mm. The number of molecules participating in the
fast magnetisation reversal ranges between $10^{18}$ and $10^{19}$
for large single crystals and assemblies of small crystals,
respectively. The typical value for the energy released in such
reversal is of the order of mJ. Consequently, the average power
released is of the order of 1 W. However, the power detected in
our experiments ranges from 3.0 $\mu$W at 2.0 K to 0.2 $\mu$W at
2.6 K. In case this radiation should be superradiance, the power
would be $P_{\rm SR} = N_m \hbar \omega \Gamma_{\rm SR}$, where
$N_m$ is the number of molecules at level $m$ contributing to
radiation emission, $\hbar \omega$ equals the energy difference
between levels $m$ and $m + 1$, and $\Gamma_{\rm SR} = N_m
\Gamma_{m}^{\rm{photons}}$. Considering that
$\Gamma_{m}^{\rm{photons}} \sim 10^{-7}$ s$^{-1}$ \cite{13,21},
and using $\omega \sim$ 100 GHz and $P_{\rm SR} \sim$ 3 $\mu$W as
measured in our experiments, the number of molecules participating
in the superradiance emission, $N_m$, is of the order of
10$^{12}$, which is less than the number of molecules populating
the excited spin levels. The requirement that all the spins have
the same transition frequencies clearly imposes a strong
restriction on the power emitted as superradiance. The fact that
the total duration of the emission of radiation, $\tau$, depends
exponentially on the temperature of the sample explains that small
variations in $T_{\rm{s}}$ produce large variations in $\tau$.

To summarise, we have found that the fast reversal of the total
magnetic moment of a set of Mn$_{12}$ single crystals is
accompanied by the emission of electromagnetic radiation of less
than 3 ms duration and a power of the order of $\mu$W at $T =$ 2.0
K.

This work has been supported by the European Commission Contract
No. IST-2001-33186 and by the Spanish Government Contract No.
MAT-2002-03144. A. Hern\'{a}ndez-M\'{\i}nguez acknowledges
financial support from Spanish MEC.

\end{document}